\documentclass[runningheads]{llncs}
\usepackage[T1]{fontenc}
\usepackage{graphicx}
\usepackage{siunitx}
\usepackage{tabularray}
\usepackage{booktabs}
\usepackage{subfigure}
\usepackage{multirow}
\usepackage[normalem]{ulem}
\usepackage{hhline}

\usepackage{amssymb}
  \usepackage{graphicx}
  \usepackage{amsmath}
  \usepackage{siunitx}
  \usepackage{bbding}

\begin{document}
\title{CodeExemplar: Example-Based Scaffolding for Introductory Programming in the GenAI Era}
\titlerunning{Example-Based Scaffolding for Introductory Programming in the GenAI Era}
\author{
Boxuan Ma
\and
Shin'ichi Konomi}
\authorrunning{Ma et al.}

\institute{Faculty of Arts and Science, Kyushu University, Fukuoka, Japan
}


%
\maketitle              % typeset the header of the contribution
\begin{abstract}

Generative AI (GenAI) can generate working code with minimal effort, creating a tension in introductory programming: students need timely help, yet direct solutions invite copying and can short-circuit reasoning. To address this, we propose example-based scaffolding, where GenAI provides scaffold examples that match a target task’s underlying reasoning pattern but differ in contexts to support analogical transfer while reducing copying. We contribute a two-dimensional taxonomy, design guidelines, and CodeExemplar, a prototype integrated with auto-graded tasks, with initial formative feedback from a classroom pilot and instructor interviews.

\end{abstract}
\section{Introduction}
For decades, programming pedagogy has relied on repeated practice as the primary route to fluency: students learn by grappling with problems, making mistakes, and iteratively refining their code \cite{denny2024prompt}. The advent of generative AI systems, however, now allow novices to obtain complete solutions with minimal effort \cite{ma2024enhancing}. While these systems can lower barriers and expand access to help, instructors increasingly worry about plagiarism, and over-reliance on AI that short-circuits students’ reasoning, weakening foundational skills and reducing engagement in productive problem solving \cite{fan2025beware}.

A central challenge, then, is not whether students can get code, but how to provide support that keeps them thinking \cite{li2025coderunner}. Prior work has explored different ways to avoid “give-the-answer” assistance. Promptly \cite{denny2024prompt} reframes practice as writing prompts that elicit correct code from AI. However, the final code is still generated by the model, which may encourage shallow engagement. In contrast, CodeRunner Agent \cite{li2025coderunner} embeds AI into an automated code assessment tool to provide context-aware hints rather than solutions, and CodeAid \cite{kazemitabaar2024codeaid} generates pseudocode to reduce direct copying. While these two approaches keep students responsible for writing the final code, the information they provide (brief hints or high-level pseudocode) can be sometimes too limited to help novices who lack a concrete reasoning template. Overall, these efforts are promising, but they still face a common tension: support must be actionable enough to help novices move forward, yet indirect enough to preserve their own reasoning.

One promising lens is analogical reasoning: using a structurally similar “source” problem to guide a new “target” problem, suggesting that well-designed scaffold examples can promote schema extraction and progress without enabling straightforward copying \cite{kazemitabaar2024codeaid}. Prior work on analogical transfer suggests that learners can benefit substantially from exposure to suitable source examples, because good sources make the underlying solution schema visible and reusable \cite{kao2022from,allen2024exploratory}. This perspective motivates a different use of generative AI in programming education. Rather than using AI to generate target-task solutions, we ask whether AI can generate scaffold examples: reference problems paired with runnable code and explanations that students can study and adapt. Crucially, an effective scaffold example should preserve the same underlying reasoning pattern as the target task while changing surface features, such as context, variable names, and input/output format. This design aims to support progress through analogical transfer: students can extract a transferable solution schema and apply it to the target, while the distance in surface features reduces the likelihood of direct copying of a target-specific answer.

In this paper, we propose an example-based scaffolding approach for introductory programming classes and make three contributions: (1) a two-dimensional taxonomy of scaffold examples to characterize pedagogical trade-offs; (2) \textsc{CodeExemplar}, a prototype that integrates scaffold examples alongside auto-graded tasks, with initial formative feedback to inform classroom deployment.

\section{Example-Based Scaffolding}

\subsection{Analogical reasoning}
Example-based scaffolding is grounded in analogical reasoning. 
Previous works \cite{gick1980analogical} describe analogical problem solving as a multi-step process: learners (i) \textbf{notice} a potentially relevant relationship between a source example and a target problem, (ii) \textbf{map} corresponding elements across the two problems, and (iii) \textbf{apply} the mapped structure from the source to construct a solution for the target. This perspective suggests that the educational value of a scaffold example depends not only on whether it appears similar to the target, but on whether it enables learners to recognize and transfer an underlying solution structure.

\subsection{Taxonomy of scaffold example}
Motivated by this analogy lens, we characterize the relationship between a target task and a reference example along two dimensions, structural similarity and surface similarity, as in prior work \cite{allen2024exploratory}.

\textbf{Structural similarity} refers to how much the example shares the target's underlying solution schema or reasoning pattern, such as control-flow schema, or nested iteration patterns.

\textbf{Surface similarity} captures similarity in context and representation (e.g., context, function signature, and input/output form), which primarily affects whether learners notice a connection, but may also increase copying risk when overly high.

Combining these dimensions, Figure~\ref{taxonomy} shows a taxonomy of reference examples that highlights different pedagogical trade-offs:

\begin{figure}[t]
\centering
\includegraphics[scale=0.35]{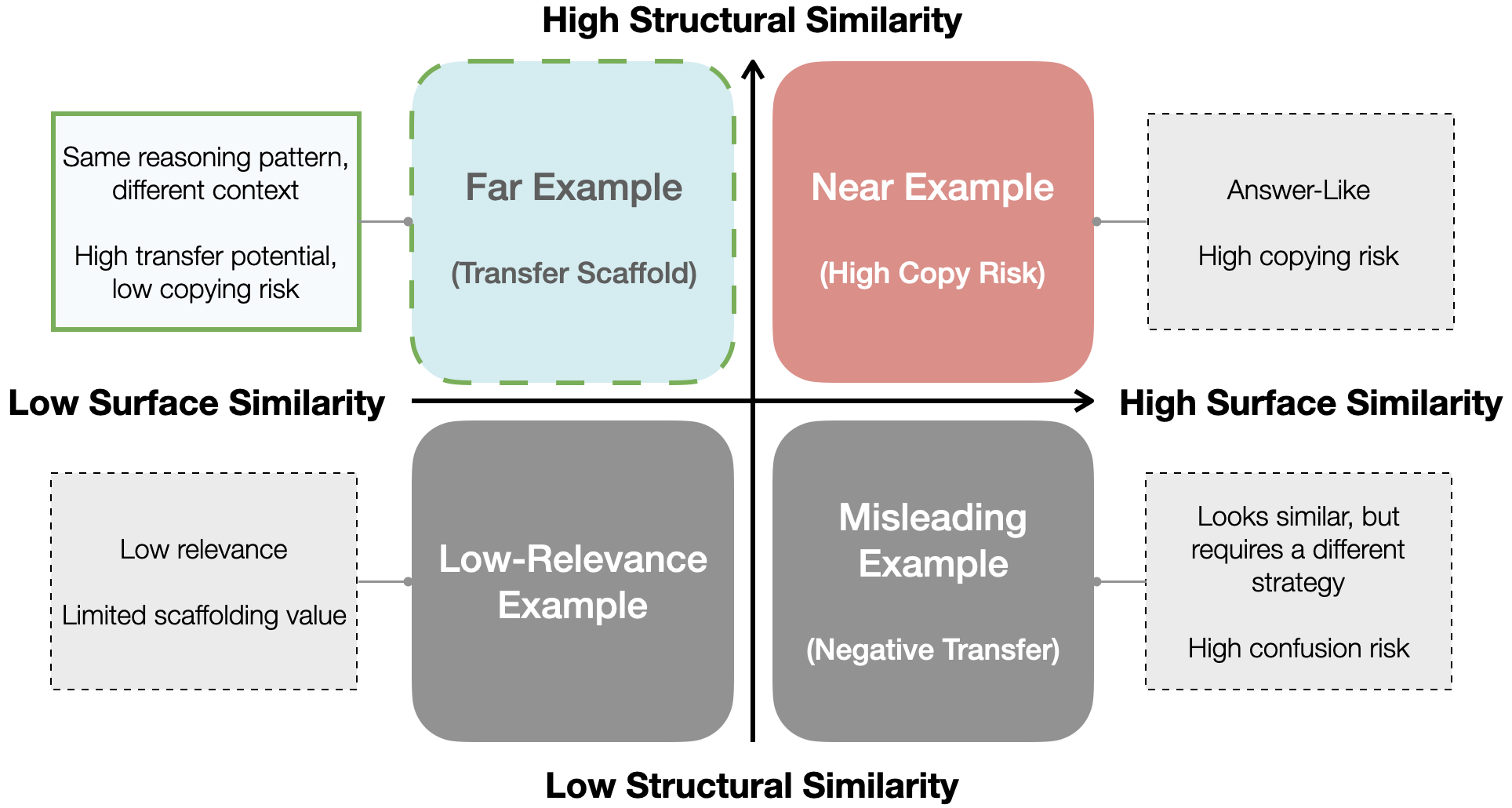}
\caption{Scaffold example taxonomy.}
\label{taxonomy}
\end{figure}

\begin{itemize}
  \item \textbf{Far Example} (\emph{high structural similarity, low surface similarity}).
  These examples preserve the same reasoning pattern as the target task while differing in surface context. They are intended to promote analogical transfer while keeping copying risk low.

  \item \textbf{Near Example} (\emph{high structural similarity, high surface similarity}).
  These examples closely resemble the target both in solution structure and appearance, making them immediately helpful but prone to being copied as a near-direct substitute for the target solution.

  \item \textbf{Misleading Example} (\emph{low structural similarity, high surface similarity}).
  These examples look similar on the surface but require a different underlying strategy. They can mislead learners and increase confusion.

  \item \textbf{Low-Relevance Example} (\emph{low structural similarity, low surface similarity}).
  These examples are distant in both structure and surface, and therefore offer limited scaffolding value for the target task.
\end{itemize}

In this work, we primarily target \emph{Far Example} as scaffold examples, aiming to support learning progress through shared reasoning patterns while discouraging direct copying.

\begin{figure}[t]
\centering
\includegraphics[scale=0.35]{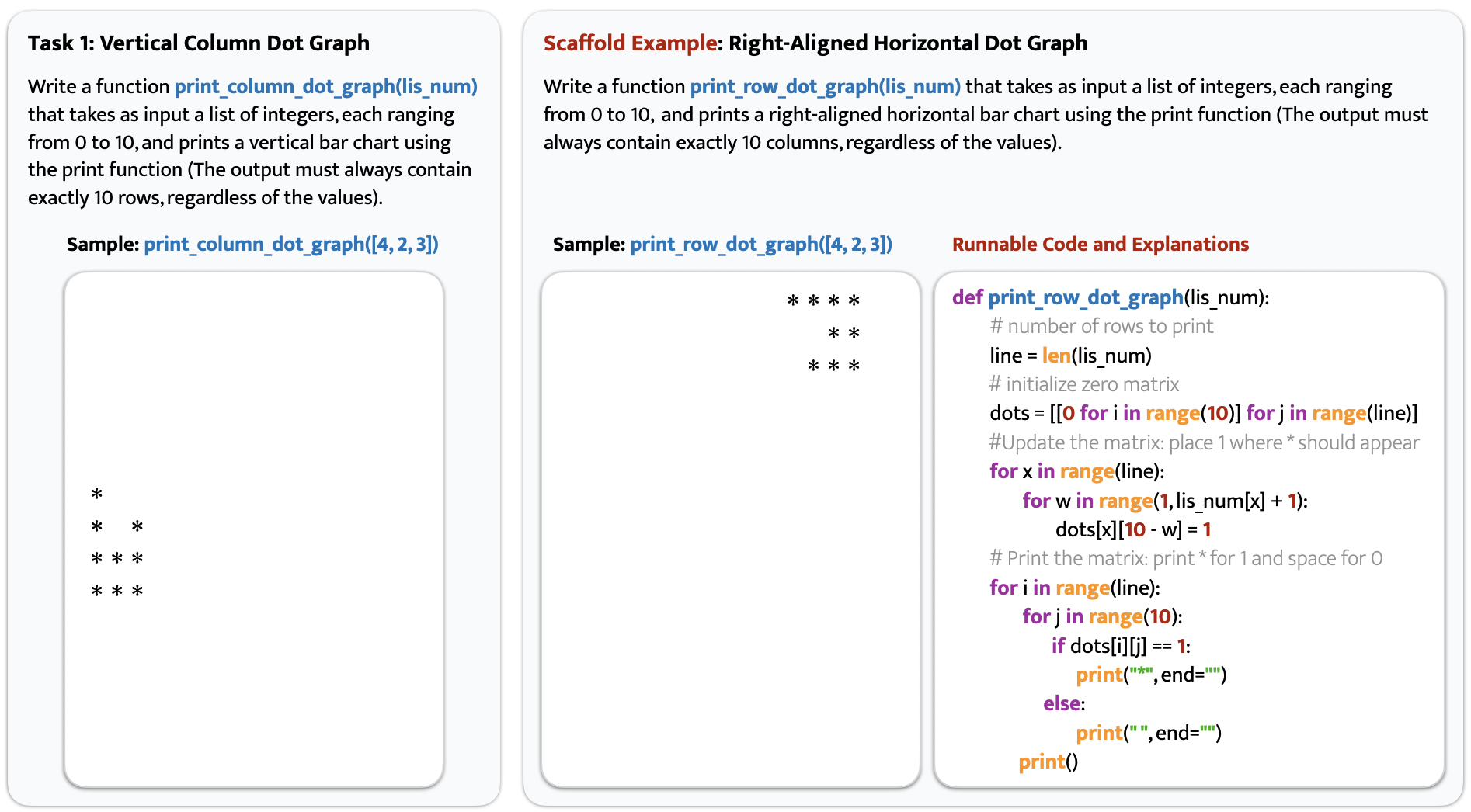}
\caption{Example from the pilot study: target task (left) and its scaffold example (right), designed to discourage direct copy-and-paste and support analogical reasoning.}
\label{example}
\end{figure}

\section{Pilot Study}

To motivate the need for example-based scaffolding and to gather early feedback from students, we conducted a small pilot study across two class sessions of an introductory programming course. The pilot involved 30 students, most of whom were first-year undergraduates.

Rather than providing solution hints to the target problems, we explored whether scaffold examples could help students make progress without simply copying an answer. We first selected three programming exercises representative of typical introductory content. For each target exercise, we used a generative AI model (GPT-5.2) to create a corresponding scaffold example, including (i) an example problem statement and (ii) runnable reference code with brief explanations. Figure~\ref{example} shows one target task and its corresponding scaffold example generated by the model. To ensure suitability for classroom use, an experienced instructor reviewed each generated scaffold for correctness and appropriateness before it was shown to students. The three target problems and their scaffold examples were then used in two class sessions. During the activity, students worked on the target problems, and were allowed to refer to the scaffold examples when they felt stuck. To gauge student perceptions of scaffold example, students in both courses were invited to fill out a survey that provided feedback on their experience.

\begin{figure}[t]
\centering
\includegraphics[scale=0.27]{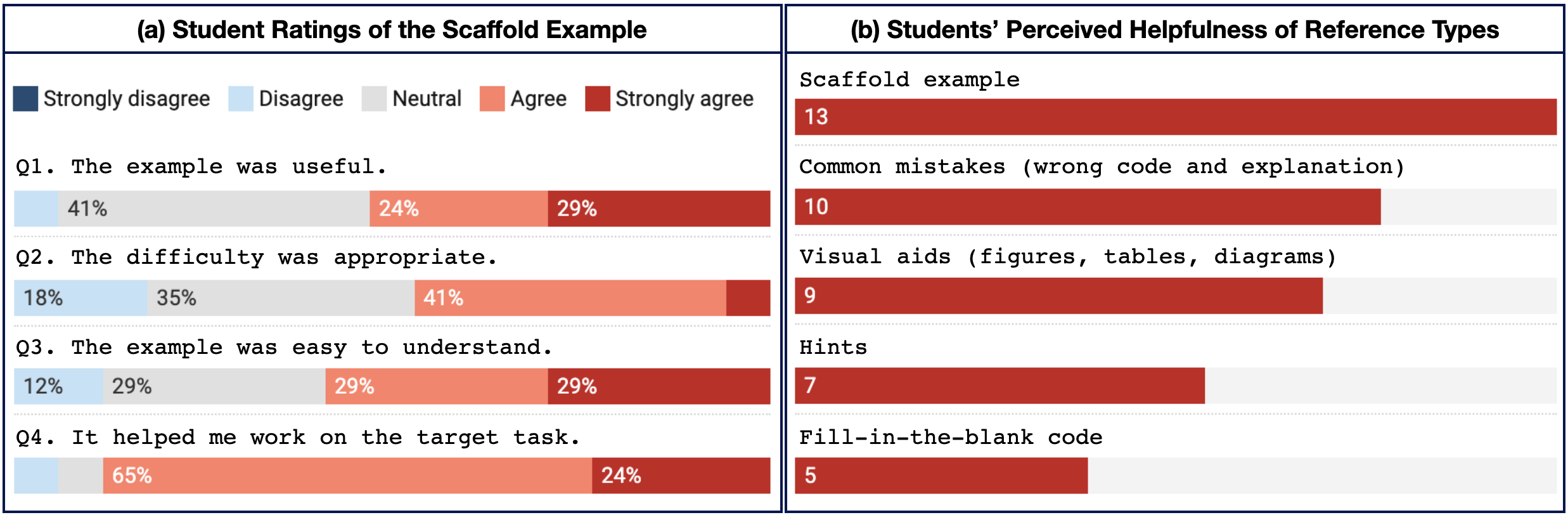}
\caption{Student ratings of scaffold examples.}
\label{pilot}
\end{figure}

A total of 17 students responded to the post-activity survey. The survey included four 5-point Likert items, one multiple-selection item, and two open-ended questions capturing overall impressions. As shown in Figure~\ref{pilot}(a), students reported positive perceptions of the scaffold examples. On a 5-point Likert scale (1 = strongly disagree, 5 = strongly agree), students rated the scaffold examples as useful overall (M = 3.76, SD = 0.97). They also rated the difficulty as appropriate (M = 3.35, SD = 0.86) and the examples as easy to understand (M = 3.76, SD = 1.03). Overall, most students agree or strongly agree that scaffold examples helped them make progress on the target task (M = 4.06, SD = 0.75). 

Figure~\ref{pilot}(b) shows which forms of support students think were helpful for them (multiple-selection), most of them chose scaffold examples, followed by common mistakes with explanations and visual aids such as figures or tables. Short hints and fill-in-the-blank code were selected less often, but were still requested by some students. 

Open-ended comments echoed these patterns: students valued scaffold examples as a way to ``get unstuck'' and to understand alternative approaches, while also requesting more adaptive or stepwise support (e.g., layered hints or optional reveal) and additional explanation of common errors. Overall, the findings suggest that scaffold examples are perceived as a promising form of support while also highlighting the need for controlled disclosure and multiple scaffold formats. Based on this feedback, we designed CodeExemplar, a prototype that generates scaffold examples on demand, with the goal of supporting learning progress without directly revealing target solutions.

\section{CodeExemplar}

\subsection{Tool Design}

\begin{figure}[t]
\centering
\includegraphics[width=\textwidth]{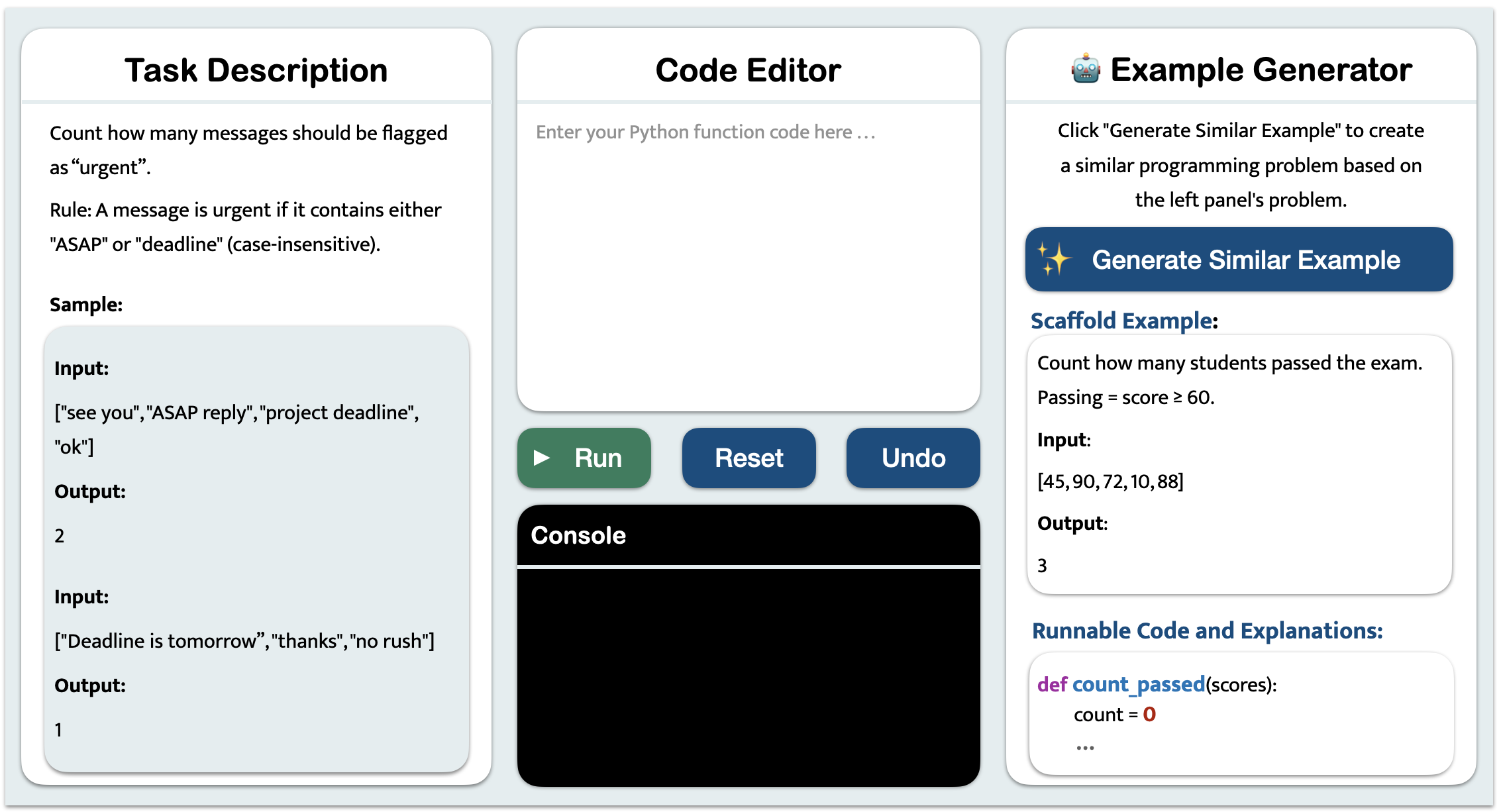}
\caption{CodeExemplar interface.}
\label{interface}
\end{figure}

Figure~\ref{interface} illustrates the interface of CodeExemplar that consists of three panels. The \textbf{Task Description} panel (left) presents the target programming task along with a small set of sample input-output pairs to clarify requirements. The \textbf{Code Editor} panel (center) is where students write and run their solution code and view outputs in a console. The \textbf{Example Generator} panel (right) provides an AI-generated \emph{scaffold example}, as well as runnable code and a brief explanation for that example.

Students first read the target task and attempt a solution in the editor. When they feel stuck, they can request a scaffold example by clicking the ‘Generate Similar Example’ button. Students are expected to use the scaffold example as a reference to infer the underlying reasoning pattern and then apply it to the target task.

\subsection{Feedback on CodeExemplar}

To obtain early feedback on the proposed design, we introduced CodeExemplar to students in class. In class, students were shown the CodeExemplar and tried it on one practice problem. They then completed a short questionnaire consisting of one overall impression item (5-point scale) and two open-ended questions. In addition, we invited four instructors who teach programming class to provide expert feedback. We presented the tool to them, allowed them to try the prototype, and collected their comments on its usefulness, potential risks, and desired features for classroom deployment.

\subsection{Student Feedback}

Students' overall impressions were favorable: 69.3\% of students reported a positive impression, including 38.5\% who reported a strongly positive impression. We also conducted a thematic analysis of students' open-ended feedback and identified the following themes.

\textbf{Actionable support} Many students emphasized that a related example could help them make progress when they were stuck, without immediately revealing the target solution. They valued that the scaffold example was concrete enough to act on (e.g., clarifying the intended reasoning steps). One student noted, \textit{``It is helpful when I don't know how to proceed. I can refer to the example and move forward,''} while another echoed \textit{``the example can provide a practical next step without giving away the answer.''}

\textbf{Comprehension support} A recurring theme was the value of pairing example code with a short explanation. Students commented that explanations made the reference easier to interpret and reduced the effort of understanding unfamiliar code. For instance, one student stated that \textit{``having both the code and an explanation makes it easier to understand what is happening,''} and another similarly highlighted that \textit{"explanations help me connect the code to the underlying idea".}

\textbf{Learning-by-analogy} Several students suggested that the example is most beneficial when it shares the same reasoning pattern as the target task while differing in surface features. This reflects the intended use of CodeExemplar as an analogical scaffold: students can notice a correspondence, map key elements, and apply the mapped strategy to the target task. As one student put it, \textit{``Even if the context is different, I can learn the core idea and reuse it for the original task.''}
 
\textbf{Concerns and feature requests} A minority of students expressed concerns that examples might reduce productive struggle or encourage dependence. These comments point to the need for learner control over how examples are revealed. For example, one student cautioned that \textit{``if examples appear too quickly, it might reduce opportunities to think on my own,''} suggesting the value of options such as delaying scaffold display, limited number of generations, or make it faded. Students also proposed several concrete improvements, such as multiple alternative scaffold examples based on request, controllable options that could adjust the scaffold (e.g., closer vs.\ farther) to better match their current understanding. Collectively, these requests informing the next iteration of our tool design.

%In addition, students wanted richer guidance beyond a single correct example, such as brief visual aids. 
\subsection{Educator Reflection}

We also gathered reflections from four instructors with experience teaching programming or computing-related courses in our university. Their teaching experience ranged from 1 to 10 years. All instructors responded positively to CodeExemplar and viewed example-based scaffolding as a promising way to support students without revealing the target solution directly. They highlighted the tool’s indirect answers as a mechanism for transfer-focused scaffolds. One instructor commented that \textit{``it forces you to think and apply the skill into practice,''} while another stressed that \textit{``it’s not enough to understand a skill—you need to apply it.''} 

One instructor suggested that the system should explain how the two tasks are related, so that students can more easily identify what is transferable and what must be adapted. This aligns with our taxonomy motivation: scaffold examples are most useful when students can map the underlying structure from the example to the target. Another instructor pointed out that even when students recognize the overall structure, they may still fail due to small variations or last-mile details. In such situations, a single example may not be sufficient, and students may require additional, more targeted support. He therefore suggested incorporating students' incorrect solutions into the support loop: the system could take a student's current attempt and errors into account and generate a more appropriate scaffold example or explanation that addresses the specific misconception.

\section{Conclusion}

In this paper, we introduced CodeExemplar, an example-based scaffolding approach that aims to support students who get stuck on introductory programming tasks without simply giving away target solutions. Our core idea is to provide scaffold examples that share an underlying reasoning pattern with the target task while differing in surface context, encouraging analogical transfer and discouraging direct copying. Future work will systematically evaluate CodeExemplar's impact on learning. In particular, we plan to investigate how such tool could influence students' progress, understanding, and reliance on AI-generated assistance.


\bibliographystyle{splncs04}
\bibliography{bib}

\end{document}